\documentclass[journal]{IEEEtran}

\usepackage{cite}
\usepackage{amsmath,amssymb,amsfonts}
\usepackage{algorithmic}
\usepackage{graphicx}
\usepackage{textcomp}

\usepackage{float}
\usepackage{color,xcolor}
\usepackage{subfigure}
\usepackage{adjustbox}
\usepackage[switch]{lineno}
\usepackage{bbding}
\usepackage{setspace}
\usepackage[labelsep=period]{caption}

\begin{document}

\title{Deformable Non-local Network for Video Super-Resolution}

\author{Hua Wang, Dewei Su, Chuangchuang Liu, Longcun Jin, Xianfang Sun, and Xinyi Peng}

\maketitle

\begin{abstract}
The video super-resolution (VSR) task aims to restore a high-resolution (HR) video frame by using its corresponding low-resolution (LR) frame and multiple neighboring frames. At present, many deep learning-based VSR methods rely on optical flow to perform frame alignment. The final recovery results will be greatly affected by the accuracy of optical flow. However, optical flow estimation cannot be completely accurate, and there are always some errors. In this paper, we propose a novel deformable non-local network (DNLN) which is a non-optical-flow-based method. Specifically, we apply the deformable convolution and improve its ability of adaptive alignment at the feature level. Furthermore, we utilize a non-local structure to capture the global correlation between the reference frame and the aligned neighboring frames, and simultaneously enhance desired fine details in the aligned frames. To reconstruct the final high-quality HR video frames, we use residual in residual dense blocks to take full advantage of the hierarchical features. Experimental results on benchmark datasets demonstrate that the proposed DNLN can achieve state-of-the-art performance on VSR task.
\end{abstract}


\section{Introduction}
\label{sec:introduction}
\IEEEPARstart{T}{he} target of super-resolution (SR) is to generate a corresponding high-resolution (HR) image or video from its low-resolution (LR) version. As an extension of single image super-resolution (SISR), video super-resolution (VSR) provides a solution to restore the correct content from the degraded video, so that the reconstructed video frames will contain more details with higher clarity. Such kind of technology with important practical significance can be widely used in many fields such as video surveillance~\cite{seibel2017eyes}, ultra-high definition television~\cite{park2018high} and so on.

Different from SISR which only considers one single low-resolution image as input at a time, VSR devotes to effectively making use of intrinsic temporal information among multiple low-resolution video frames. Although vanilla SISR approaches can be directly applied to video frames by treating them as single images, abundant detail information available from neighboring frames will be wasted. Such practice is hard to reconstruct promising video frames, so they are not well adapted to VSR task.

To overcome the limitation of the SISR, existing VSR methods~\cite{park2003super,farsiu2004fast,liu2013bayesian,ma2015handling,wang2018video} usually take a LR reference frame and its multiple neighboring frames as inputs to reconstruct a corresponding HR reference frame. Due to the motion of the camera or objects, the neighboring frames should be spatially aligned first for utilizing the temporal information. To this end, most traditional VSR methods~\cite{liu2017robust,tao2017detail,sajjadi2018frame,caballero2017real} generally calculate the optical flow and estimate the sub-pixel motion between LR frames to achieve the alignment operation. However, fast and reliable flow estimation still remains a challenging problem. The brightness constancy assumption, which most motion estimation algorithms rely on, may be invalid due to the existence of motion blur and occlusion. Incorrect motion compensation will introduce artifacts in aligned neighboring frames and affect the quality of final reconstructed video frames. Hence, explicit flow estimation and motion compensation methods could be sub-optimal for VSR task. 

In this paper, we propose a novel deformable non-local network (DNLN), which is non-optical-flow-based, to perform both implicit motion estimation and video super-resolution. Our network mainly consists of four modules: feature extraction module, alignment module, non-local attention module and SR reconstruction module. Inspired by TDAN~\cite{tian2018tdan}, we apply the deformable convolution~\cite{zhu2019deformable} in our alignment module and enhance its ability of adaptively warping frames. Specifically, we introduce a hierarchical feature fusion block (HFFB)~\cite{hui2019progressive} to effectively handle the videos with large and complex motions. Through the stacks of deformable convolutions, we align the neighboring frame to the reference frame at the feature level and gradually improve the alignment accuracy. Then in the non-local attention module, we exploit a non-local structure to capture the global correlation between the reference feature and each aligned neighboring feature, which assesses the importance of different regions in neighboring feature. Such operation is expected to highlight the features complementary to the reference frame and exclude regions with improper alignment. The features with attention guidance are fused and then fed into the final SR reconstruction module. Here, we use residual in residual dense blocks (RRDB)~\cite{wang2018esrgan} to generate the SR reference frame. RRDBs help to make full use of the information from different hierarchical levels and retain more details of the input LR frame.

In summary, the main contributions of this paper can be concluded as follows: 
\begin{itemize}
	
	\item We propose a novel deformable non-local network (DNLN) to accomplish high quality video super-resolution. Our method achieves the most advanced VSR performance on several benchmark datasets.
	
	\item We design an alignment module based on deformable convolution, which can realize the feature level alignment in a coarse to fine manner without explicitly motion compensation.
	
	\item We propose a non-local attention module to select significant features from neighboring frames which are conducive to the recovery of the reference frame.
	
\end{itemize}

The rest of the paper is organized as follows: Section 2 introduces the related work. Section 3 elaborates the structure of the proposed network. Section 4 shows our experimental results on benchmark datasets, including visual comparisons with other methods. The effectiveness of each components in our network is analyzed in Section 5. Finally, Section 6 draws conclusions.

\section{Related Work}
\subsection{Single Image Super-resolution}

Dong et al.~\cite{dong2014learning} first proposed SRCNN for single image super-resolution to learn the nonlinear mapping between LR and HR images in an end-to-end manner, which achieves better performance than previous work. Kim et al.~\cite{kim2016accurate} further improved SRCNN by stacking more convolution layers and using residual learning to increase network depth. Tai et al. introduced recursive blocks in DRRN~\cite{tai2017image}, which employs parameters sharing strategy to make the training stable. All of these methods first upscale the LR input to the desired size and the reconstruction process is based on the upscaled products. Such pre-processing step inevitably results in loss of details and additional computation cost. To avoid these problems, extracting features from the original LR input and upscaling spatial resolution at the end of the network become the main direction of SR network. Dong et al.~\cite{dong2016accelerating} directly took the original LR image as input and brought in the transpose convolution layer (also known as the deconvolution layer) for upsampling features to high resolution outcomes. Shi et al.~\cite{shi2016real} proposed an effective sub-pixel convolution layer for amplifying the final LR feature map to SR output and accelerating the network.

Afterwards, Timofte et al.~\cite{timofte2017ntire} provided a new large dataset (DIV2K) in the NTIRE 2017 challenge that consists of 1000 2K resolution images. This dataset enables researchers to train deeper and wider networks which leads to various development of SR methods. The most advanced SISR networks, such as EDSR~\cite{lim2017enhanced}, DBPN~\cite{haris2018deep}, RDN~\cite{zhang2018residual} and RCAN~\cite{zhang2018image}, have far better training performance on this dataset than previous networks.

\subsection{Video Super-resolution}

Liao et al.~\cite{liao2015video} proposed DECN and made use of two classical optical flow methods: TV-L1 and MDP flow to generate SR drafts with different parameters, and then produced the final result through a deep network. Kappeler et al.~\cite{kappeler2016video} proposed VSRnet, which uses a hand-designed optical flow algorithm to perform motion compensation on the input LR frame, and takes the warped frame as the CNN input to predict the HR video frame. Caballero et al.~\cite{caballero2017real} introduced the first end-to-end VSR network: ESPCN, which studies early fusion, slow fusion, and 3D convolution to learn temporal relationships. They applied a multi-scale spatial transformer to warp the LR frame and eventually generated a HR frame through another deep network. Tao et al.~\cite{tao2017detail} proposed a sub-pixel motion compensation layer for frame alignment and used a convolutional LSTM architecture in following SR reconstruction network. Recently, Haris et al.~\cite{haris2019recurrent} proposed RBPN which learns from the idea of back-projection to iteratively extract temporal features between frames. They treated frames independently rather than concatenated them together.

\begin{figure*}[t!]
	\centering
	\includegraphics[width=\linewidth]{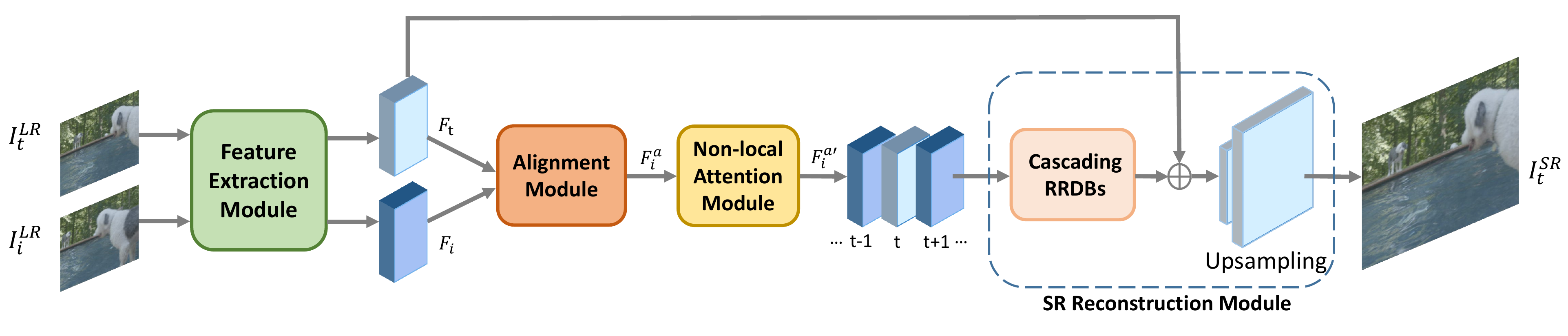}
	\caption{The architecture of the proposed DNLN framework. We only show one neighboring frame in the figure. Each neighboring frame will pass through feature extraction module, alignment module and non-local attention module. Then all the features are concatenated and fed into SR reconstruction module to generate HR reference frame.}
	\label{figure:proposed_network}
\end{figure*}

Most previous VSR methods exploit optical flow to estimate motion between frames and perform motion compensation to integrate effective features. While various approaches~\cite{dosovitskiy2015flownet,ranjan2017optical,ilg2017flownet,hui2018liteflownet,zhai2019skflow} are proposed to calculate the optical flow, it is still intractable to obtain precise flow estimation in the case of occlusion and large movement. Xue et al.~\cite{xue2017video} proposed task-oriented TOFlow with learnable task-oriented motion prompts. It achieved better VSR results than fixed flow algorithm, which reveals that standard optical flow is not the best motion representation for video recovery. To circumvent this problem, DUF~\cite{jo2018deep} uses an adaptive upsampling with dynamic filters instead of the explicit estimation process. TDAN~\cite{tian2018tdan} uses deformable convolutions to adaptively align the video frame at the feature level without computing optical flow. These kind of methods transcend the flow-based approaches through implicit motion compensation.

\subsection{Deformable Convolution}

 To enhance the CNNs’ capability of modeling geometric transformations, Dai et al.~\cite{dai2017deformable} proposed deformable convolutions. It adds additional offsets to the regular grid sampling locations in the standard convolution and enables arbitrary deformation of the sampling grid. To further enhance the modeling capability, they proposed modulated deformable convolutions~\cite{zhu2019deformable} which can further learn modulation scalar for sampling kernels. The modulation scalar lies in the range $[ 0,1 ]$, which can adjust the weight for each sampling location. The deformable convolution is effective for high-level vision tasks such as object detection and semantic segmentation. TDAN~\cite{tian2018tdan} is the first to utilize deformable convolutions in the VSR task. It is an end-to-end network which adaptively aligns the input frames at the feature level without explicit motion estimation. EDVR~\cite{wang2019edvr} further exploits the deformable convolutions with a pyramid and cascading structure. It shows superior performance to previous optical-flow-based VSR networks.

\subsection{Non-local Block}

Inspired by the classic non-local method in computer vision, Wang et al.~\cite{wang2018non} proposed a building block for video classification by virtue of non-local operations. For image data, long-range dependencies are commonly modeled via large receptive fields formed by deep stacks of convolutional layers. While the non-local operations capture long-range dependencies directly by computing interactions between any two positions, regardless of their positional distance. It computes the response at a position as a weighted sum of all positions in the input feature maps. The set of positions can be in space, time, or spacetime, so the non-local operations are applicable for image or video problems.

\section{Deformable Non-local Networks}

\subsection{Network Architecture}

Given a sequence of 2N+1 consecutive low-resolution frames$\left\{ I _ { t - N } ^ { LR } , \ldots , I _ { t - 1 } ^ { LR } , I _ { t } ^ { LR } , I _ { t + 1 } ^ { LR } , \ldots , I _ { t + N } ^ { LR } \right\}$, where $I _ { t } ^ { LR }$ is the reference frame and the others are the neighboring frames, our goal is to recover the corresponding high quality video frame through the reference frame and its 2N neighboring frames. Therefore, our network takes $I _ { [ t - N , t + N ] } ^ { LR }$ as inputs, and finally reconstructs $I _ { t } ^ { SR }$. The overall network structure is shown in Fig.\ref{figure:proposed_network}, which can be divided into four parts, including feature extraction module, alignment module, non-local attention module and the final SR reconstruction module.

For all the input LR frames, we first extract their features via a shared feature extraction module. It consists of one convolutional layer and several residual blocks. The feature extraction can be represented as: 
\begin{equation}
F_{T}=H_{fea}\left(I_{T}^{L R}\right),
\end{equation}
where the output $F_{T}$ denotes the extracted LR feature maps. Then each LR neighboring feature $F_{i}$ will enter the alignment module along with the LR reference feature $F_{t}$. Our alignment module which consists of stacked deformable convolutions is responsible for performing adaptive feature level alignment: 
\begin{equation}
F_{i}^{a}=H_{align}\left(F_{i}, F_{t}\right),
i \in[t-N, t+N] \text { and } i \neq t,
\end{equation}
where $F_{i}^{a}$ denotes the neighboring feature after alignment. Subsequently, each aligned neighboring feature and the reference feature are fed into a non-local attention module. By calculating the global correlation between them, connections of pixels are established and informative regions in $F_{i}^{a}$ will be further enhanced. The output $F_{i}^{a'}$ of the non-local attention module can be expressed as:
\begin{equation}
F_{i}^{a'}=H_{nl}\left(F_{i}^{a}, F_{t}\right).
\end{equation}
The last part is the SR reconstruction module, here we use the residual in residual dense blocks (RRDB). We concatenate 2N+1 features and fuse them through a $3 \times 3$ convolution layer, then the fused feature maps are fed into RRDBs for further reconstruction. Besides, we use a skip connection to propagate LR reference feature to the end of the network and do an element-wise addition with the outcome of RRDBs. Finally, a high quality HR reference frame is recovered from the output feature. The reconstruction module is defined as follows:
\begin{equation}
F_{\text {fusion}}=\operatorname{Conv}\left(\left[F_{t-N}^{a'}, \ldots, F_{t-1}^{a'}, F_{t}, F_{t+1}^{a'}, \ldots, F_{t+\mathrm{N}}^{a'}\right]\right),
\end{equation}
\begin{equation}
I_{t}^{S R}=H_{r e c}\left(H_{R R D B s}\left(F_{f u s i o n}\right)+F_{t}\right),
\end{equation}
where $[\cdot, \cdot, \cdot]$ denotes concatenation of the features. $H_{r e c}$ contains an upscaling layer and a reconstruction layer.

\begin{figure*}[t!]
	\centering
	\includegraphics[width=\linewidth]{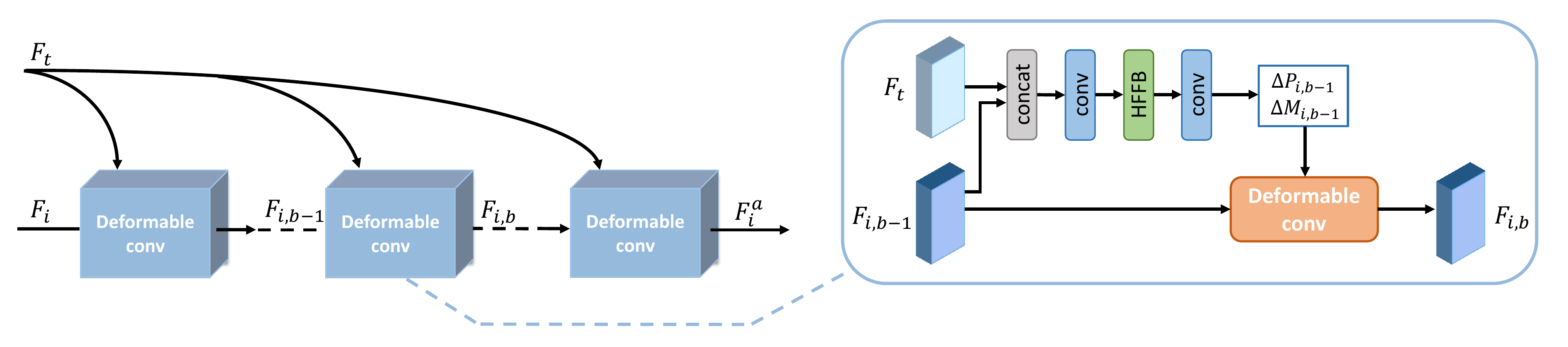}
	\caption{The proposed alignment module and the detailed illustration of deformable convolution operation.}
	\label{figure:alignment module}
\end{figure*}

\begin{figure*}[htpb]
	\begin{center}
		\subfigure[Hierarchical feature fusion block (HFFB)]
			{\includegraphics[width=0.45\textwidth]{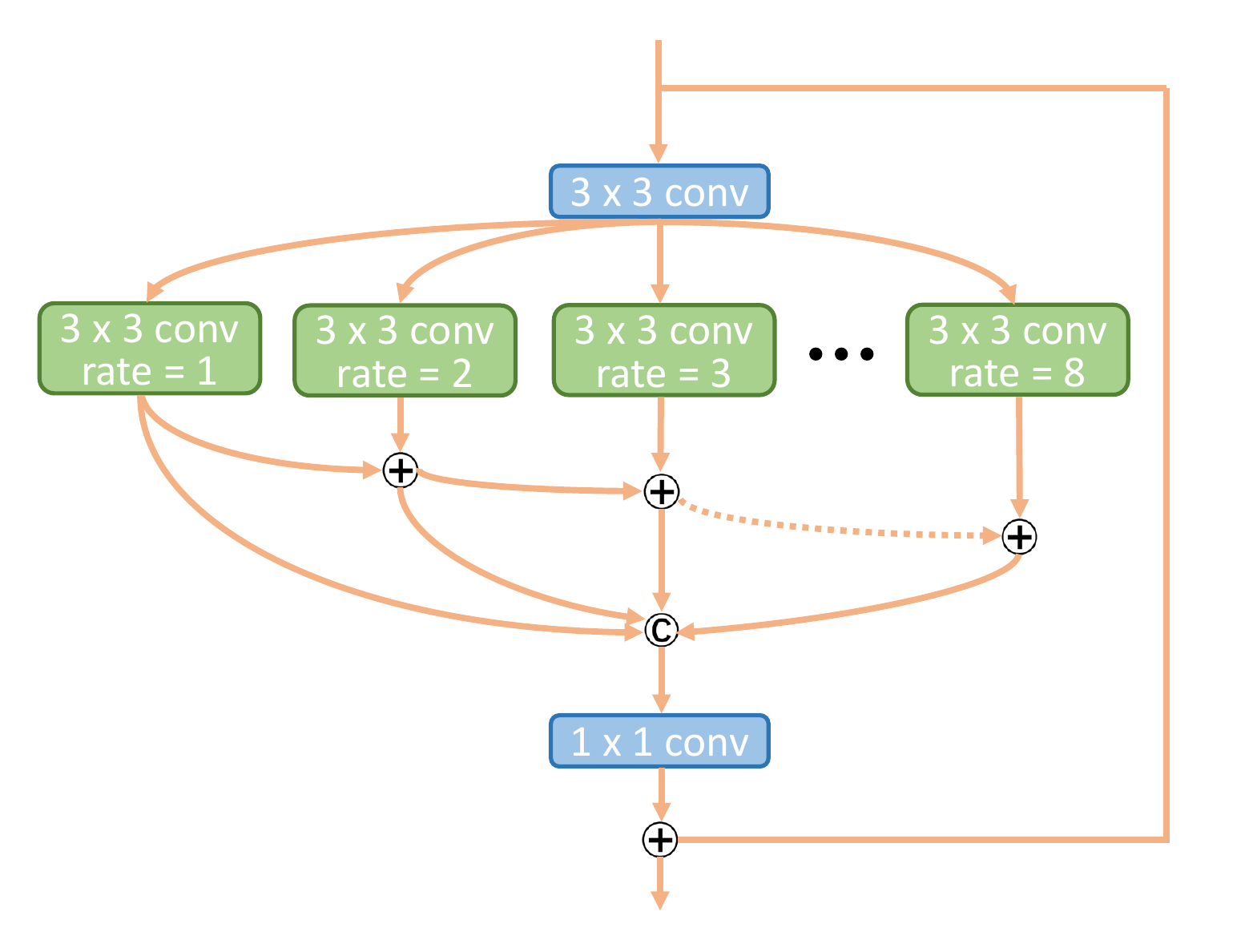}}
		\hfil
		\subfigure[Receptive field of multiple dilated convolutions addition]
			{\includegraphics[width=0.45\textwidth]{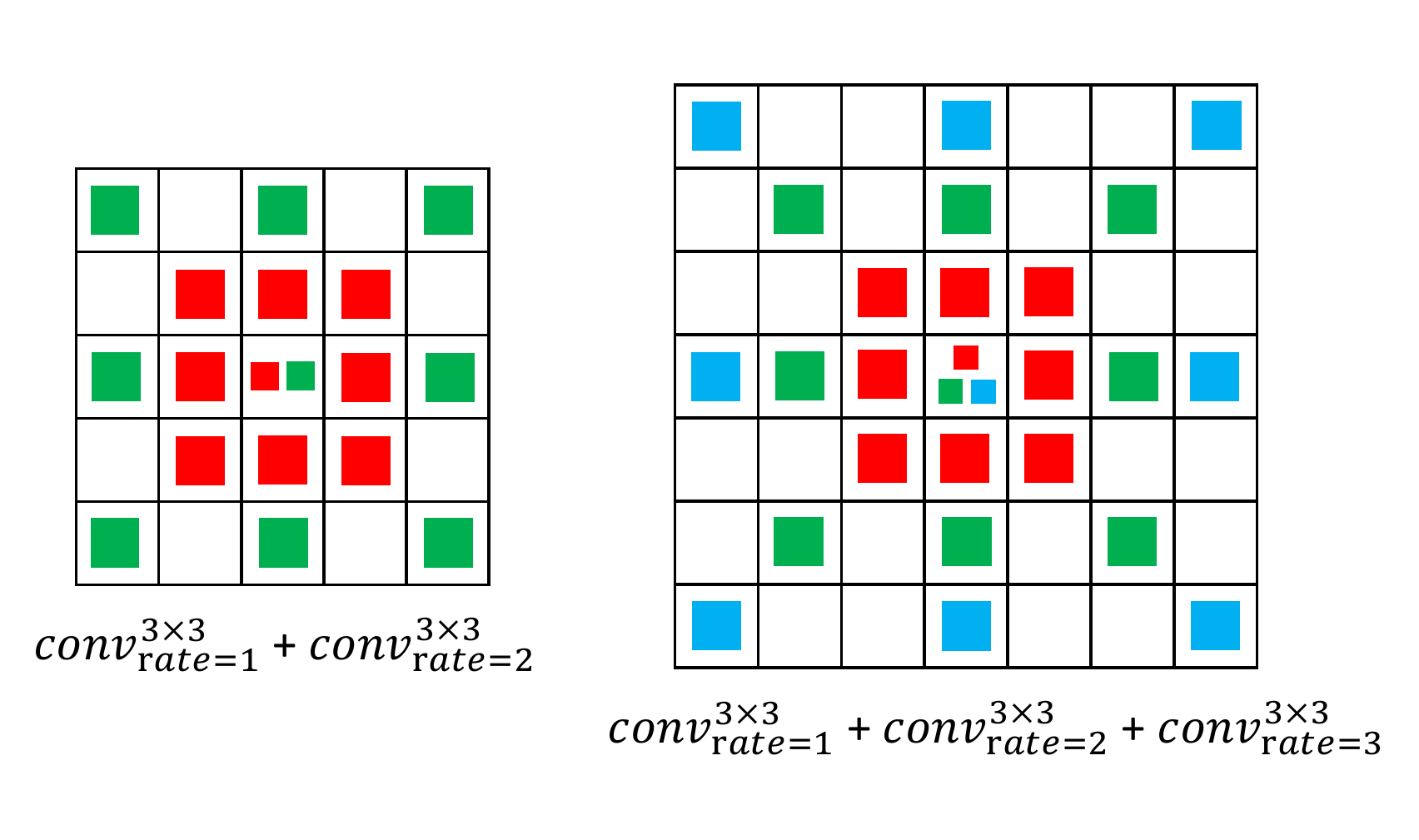}}
		\caption{(a) The structure of hierarchical feature fusion block (HFFB). It contains 8 $3 \times 3$ dilated convolutions with a dilation rate from 1 to 8. The feature maps obtained using kernels of different dilation rates are hierarchically added before being concatenated. (b) A diagrammatic sketch of multiple dilated convolutions addition.}
		\label{figure:hffb}
	\end{center}
\end{figure*}

\subsection{Alignment Module}
In order to make use of temporal information from consecutive frames, traditional VSR methods are based on optical flow to perform frame alignment. However, explicit motion compensation method could be sub-optimal for video super-resolution task. We use modulated deformable convolutions~\cite{zhu2019deformable} in the alignment module to get rid of such limitation. 

For each location $p$ on the output feature map $Y$, a normal convolution process can be expressed as:
\begin{equation}
Y(p)=\sum_{k=1}^{K} \omega_{k} \cdot X\left(p+p_{k}\right),
\end{equation}
where $p_{k}$ represents the sampling grid with $K$ sampling locations and $\omega_{k}$ denotes the weights for each location. For example, $K=9$ and $p_{k} \in\{(-1,-1),(-1,0), \ldots,(1,1)\}$ defines a $3 \times 3$ convolutional kernel. In the modulated deformable convolution, predicted offsets and modulation scalar are added to the sampling grid making deformable kernels spatially-variant. Here, we utilize the deformable convolution for temporal alignment. Let $F_{i,b-1}$ and $F_{i,b}$ denote the input and output of the deformable convolution in our module, respectively. The operation of modulated deformable convolution is as follows:
\begin{equation}
F_{i,b}(p)=\sum_{k=1}^{K} \omega_{k} \cdot F_{i,b-1}\left(p+p_{k}+\Delta p_{i,k}\right) \cdot \Delta m_{i,k},
\label{eq:deform_conv}
\end{equation}
where $\Delta p_{i,k}$ and $\Delta m_{i,k}$ are the learnable offset and modulation scalar for the $k$-th location, respectively. The convolution will be operated on the irregular positions with dynamic weights to achieve adaptive sampling on input features. Since the offsets and modulation scalar are both learned, each input neighboring feature will be concatenated with the reference one to generate the corresponding deformable sampling parameters:
\begin{equation}
\Delta P_{i}, \Delta M_{i}=f\left(\left[F_{i}, F_{t}\right]\right), 
\end{equation}
where $[ \cdot , \cdot ]$ denotes the concatenation operation. And $\Delta P=\{\Delta p_{k}\}$, $\Delta M=\{\Delta m_{k}\}$. As the $\Delta p_{k}$ may be fractional, we use the bilinear interpolation, which is the same as that proposed in \cite{dai2017deformable}.

The alignment module proposed in DNLN is composed of several deformable convolutions as shown in Fig.\ref{figure:alignment module}. In each deformable convolution, a reference feature $F_{t}$ and a neighboring feature $F_{i}$ are concatenated as an input. Then they pass through a $3 \times 3$ convolution layer to reduce channels and a hierarchical feature fusion block (HFFB)~\cite{hui2019progressive} to increase the size of receptive field. The following $3 \times 3$ convolution layer is used to obtain the offset $\Delta P_{i}$ and modulation scalar $\Delta M_{i}$ for the deformable kernel. The structure of HFFB is depicted in Fig.\ref{figure:hffb}. It introduces a spatial pyramid of dilated convolutions to effectively enlarge receptive field with relatively low computational cost, which contributes to deal with complicated and large motions between frames. In HFFB, the feature maps obtained using kernels of different dilation rates are hierarchically added before being concatenated. With the same size of receptive field, the multiple dilated convolutions addition is more dense than just one dilated convolution. The use of HFFB is beneficial to acquire an effective receptive field, so we can more efficiently exploit the temporal dependency of pixels to generate the sampling parameters.

According to Eq.(\ref{eq:deform_conv}), the deformable kernel can adaptively select sampling positions on neighboring features, learn implicit motion compensation between two frames, and complete the alignment of features. With a cascade of deformable convolutions, we can gradually align the neighboring features and improve the alignment accuracy of sub-pixels. It is noticed that when passing through a deformable convolution layer, the reference feature always keeps unchanged, only to provide a reference for the alignment of neighboring features. Through such a coarse to fine process, the neighboring frames can be well warped at the feature level.

\begin{figure}[t!]
	\centering
	\includegraphics[width=\linewidth]{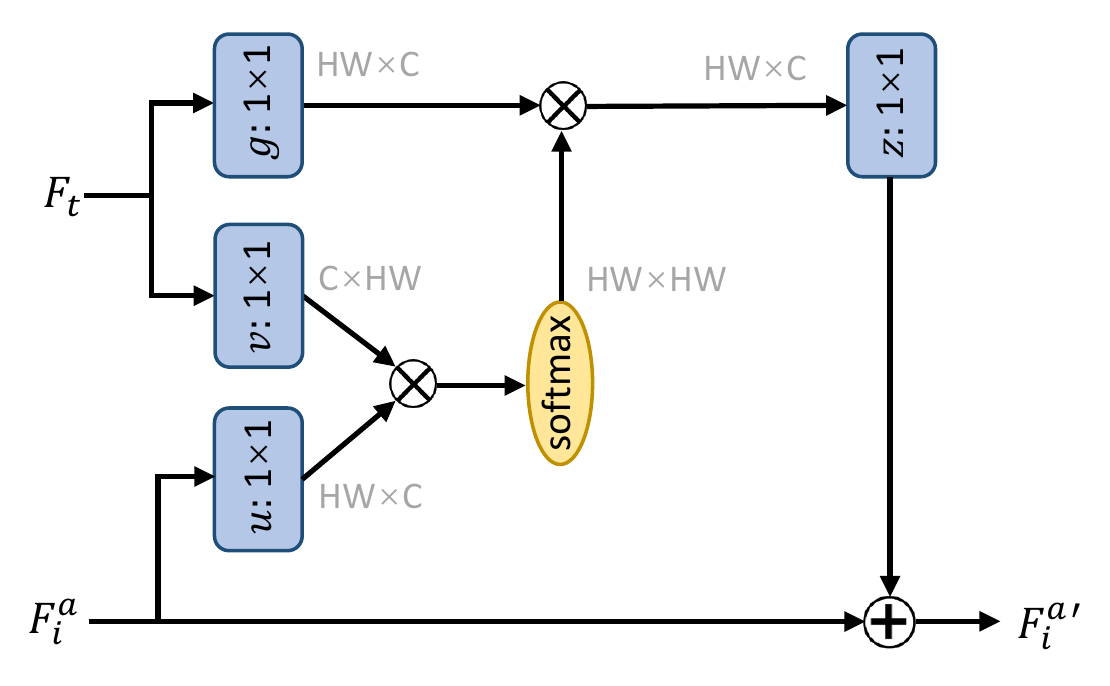}
	\caption{The non-local attention module.}
	\label{figure:non-local attention module}
\end{figure}

\subsection{Non-local Attention Module}

Due to the factors such as occlusion, blurring and parallax problems, even after the alignment module, the neighboring frames still have some areas that are not well aligned or don't contain the missing details needed for the reference frame. Therefore, it is essential to dynamically select valid inter-frame information before merging features. The proposed DNLN introduces a non-local attention module to achieve this goal. By capturing the global correlation between the aligned neighboring feature $F_{i}^{a}$ and the reference one $F_{t}$, the non-local module can effectively enhance desirable fine details in $F_{i}^{a}$ which can be complementary to the reference frame, and suppress the misaligned areas.

Let $\mathbf{x}$ and $\mathbf{y}$ denote the input feature $F_{i}^{a}$ and $F_{t}$ in Fig.\ref{figure:non-local attention module}, respectively. The non-local operation in our module can be defined as:
\begin{equation}
\mathbf{z}_{p}=\mathbf{x}_{p}+W_{z} \sum_{n=1}^{N} \frac{f\left(\mathbf{x}_{p}, \mathbf{y}_{n}\right)}{\mathcal{C}(\mathbf{y})} \left(W_{g} \cdot \mathbf{y}_{n}\right),
\end{equation}
where $\mathbf{z}$ denotes the module output $F_{i}^{a'}$. Here $p$ is the index of an output position, and $n$ is the index that enumerates all positions on $\mathbf{y}$. $W_{g} \mathbf{y}_{n}$ computes the expression of input $\mathbf{y}$ at position $n$. The function $f\left(\mathbf{x}_{p}, \mathbf{y}_{n}\right)$ calculates the relationship between $\mathbf{x}_{p}$ and $\mathbf{y}_{n}$. We use embedded Gaussian function to represent this pairwise relationship  and it is normalized by a factor $\mathcal{C}(\mathbf{y})$:
\begin{equation}
\frac{f\left(\mathbf{x}_{p}, \mathbf{y}_{n}\right)}{\mathcal{C}(\mathbf{y})}=\frac{\exp \left(\left\langle W_{u} \mathbf{x}_{p}, W_{v} \mathbf{y}_{n}\right\rangle\right)}{\sum_{n} \exp \left(\left\langle W_{u} \mathbf{x}_{p}, W_{v} \mathbf{y}_{n}\right\rangle\right)}.
\end{equation}
$W_{u} \mathbf{x}_{p}$, $W_{v} \mathbf{y}_{n}$ are used to linearly embed the input and pairwise relationship is obtained from such a softmax computation. Then we calculate a value of position $p$ by using these relationships and the corresponding expression of all positions on $\mathbf{y}$. The value is added to the input $\mathbf{x}_{p}$ to get the final output $\mathbf{z}_{p}$. Through non-local operation, the neighboring features can make full use of the correlation with the reference feature at the pixel level and enhance the desired missing details.

\begin{figure}[t!]
	\centering
	\includegraphics[width=\linewidth]{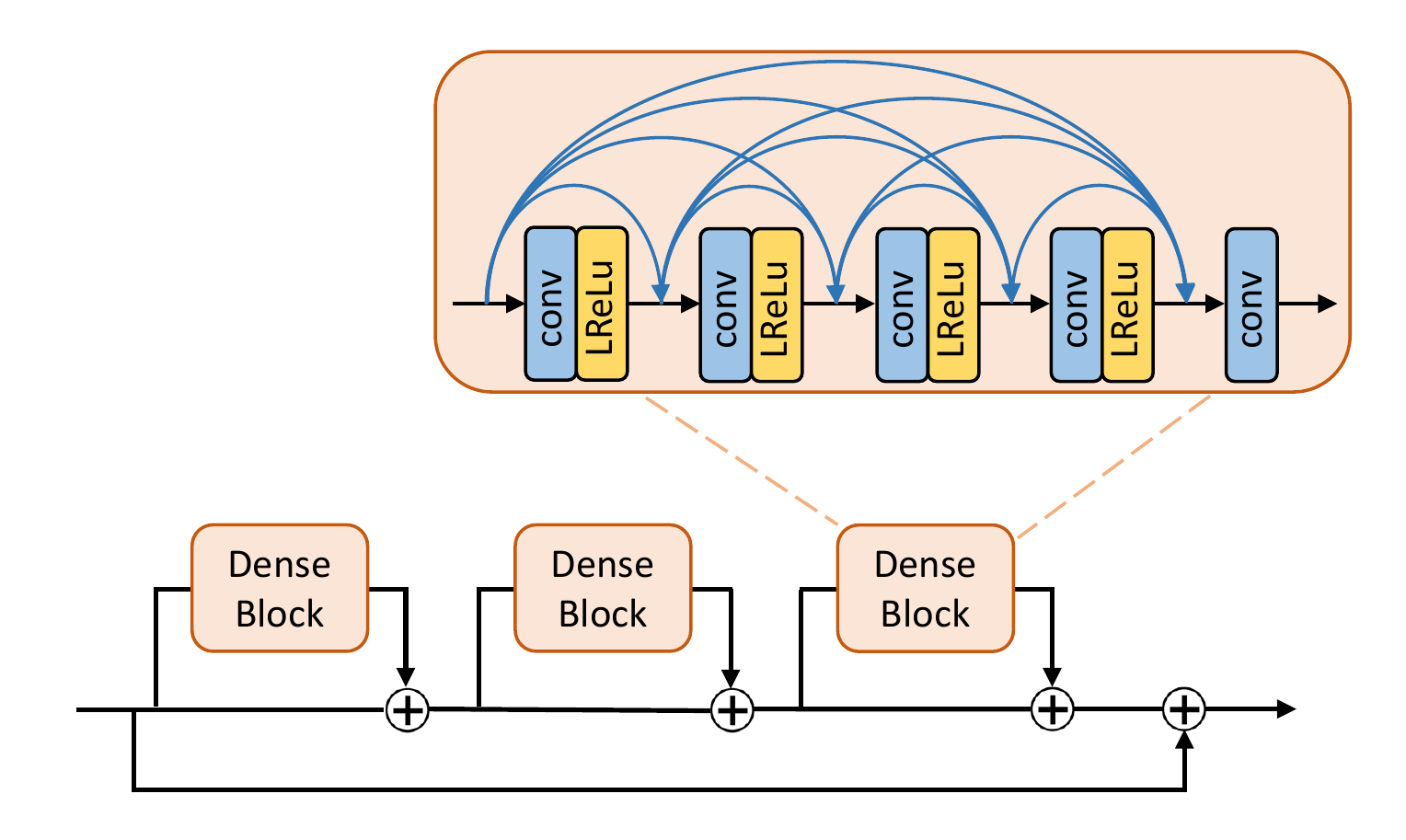}
	\caption{The residual in residual dense block (RRDB).}
	\label{figure:rrdb}
\end{figure}


\begin{figure*}[t!]
	\begin{center}

    \begin{tabular}{cc}
		\begin{adjustbox}{valign=t}
		\tiny
			\begin{tabular}{c}
				\includegraphics[width=.33\textwidth]{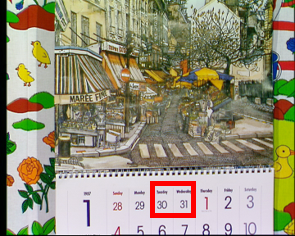}\vspace{0.5em}
				\\
				
				{\small ``Calendar'' \hspace{-1em}}\vspace{0.5em}
			\end{tabular}
		\end{adjustbox}
		\hspace{-0.4cm}
		\begin{adjustbox}{valign=t}
		\tiny
			\begin{tabular}{ccccc}
			    \centering
				\includegraphics[width=.145\textwidth,height=2.1cm]{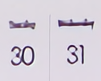}\hspace{-1em} &
				\includegraphics[width=.145\textwidth,height=2.1cm]{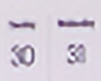} \hspace{-1em} &
				\includegraphics[width=.145\textwidth,height=2.1cm]{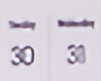} \hspace{-1em} &
				\includegraphics[width=.145\textwidth,height=2.1cm]{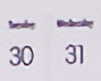}\vspace{0.42em}
				\\
				{\small (a) RCAN\hspace{-1em}}
			    &{\small (b) VESPCN \hspace{-1em}}
			    &{\small (c) TOFlow\hspace{-1em}}
		    	&{\small (d) FRVSR\hspace{-1em}}\vspace{0.42em}
				\\
			
				\includegraphics[width=.145\textwidth,height=2.1cm]{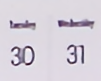} \hspace{-1em} &
				\includegraphics[width=.145\textwidth,height=2.1cm]{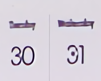} \hspace{-1em} &
				\includegraphics[width=.145\textwidth,height=2.1cm]{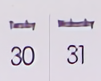} \hspace{-1em} &
				\includegraphics[width=.145\textwidth,height=2.1cm]{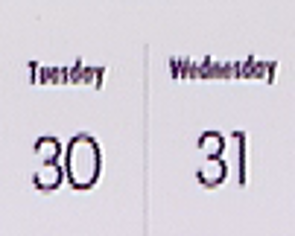}\vspace{0.5em}
				\\
				{\small (e) DUF\hspace{-1em}}
			    &{\small (f) RBPN\hspace{-1em}}
			    &{\small (g) Ours \hspace{-1em}}
			    &{\small (h) GT\hspace{-1em}}\vspace{0.5em}\\

			\end{tabular}
		\end{adjustbox}
		\vspace{0.5mm}
		\\
        \end{tabular}

    \begin{tabular}{cc}
		\begin{adjustbox}{valign=t}
		\tiny
			\begin{tabular}{c}
				\includegraphics[width=.33\textwidth]{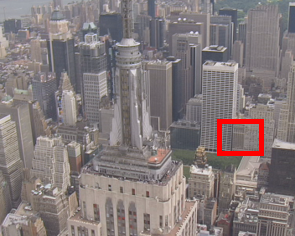}\vspace{0.5em}
				\\
				
				{\small ``City'' \hspace{-1em}}\vspace{0.5em}
			\end{tabular}
		\end{adjustbox}
		\hspace{-0.4cm}
		\begin{adjustbox}{valign=t}
		\tiny
			\begin{tabular}{ccccc}
			    \centering
				\includegraphics[width=.145\textwidth,height=2.1cm]{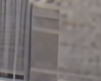}\hspace{-1em} &
				\includegraphics[width=.145\textwidth,height=2.1cm]{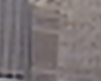} \hspace{-1em} &
				\includegraphics[width=.145\textwidth,height=2.1cm]{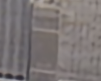} \hspace{-1em} &
				\includegraphics[width=.145\textwidth,height=2.1cm]{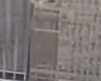}\vspace{0.42em}
				\\
				{\small (a) RCAN\hspace{-1em}}
			    &{\small (b) VESPCN \hspace{-1em}}
			    &{\small (c) TOFlow\hspace{-1em}}
		    	&{\small (d) FRVSR\hspace{-1em}}\vspace{0.42em}
				\\
			
				\includegraphics[width=.145\textwidth,height=2.1cm]{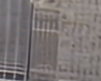} \hspace{-1em} &
				\includegraphics[width=.145\textwidth,height=2.1cm]{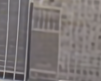} \hspace{-1em} &
				\includegraphics[width=.145\textwidth,height=2.1cm]{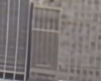} \hspace{-1em} &
				\includegraphics[width=.145\textwidth,height=2.1cm]{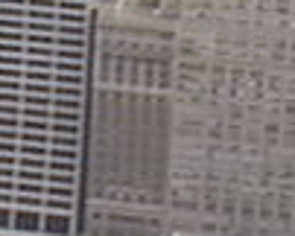}\vspace{0.5em}
				\\
				{\small (e) DUF\hspace{-1em}}
			    &{\small (f) RBPN\hspace{-1em}}
			    &{\small (g) Ours \hspace{-1em}}
			    &{\small (h) GT\hspace{-1em}}\vspace{0.5em}\\

			\end{tabular}
		\end{adjustbox}
		\vspace{0.5mm}
		\\
        \end{tabular}

    \begin{tabular}{cc}
		\begin{adjustbox}{valign=t}
		\tiny
			\begin{tabular}{c}
				\includegraphics[width=.33\textwidth]{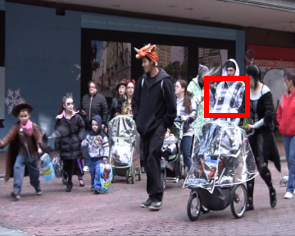}\vspace{0.5em}
				\\
				
				{\small ``Walk'' \hspace{-1em}}\vspace{0.5em}
			\end{tabular}
		\end{adjustbox}
		\hspace{-0.4cm}
		\begin{adjustbox}{valign=t}
		\tiny
			\begin{tabular}{ccccc}
			    \centering
				\includegraphics[width=.145\textwidth,height=2.1cm]{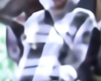}\hspace{-1em} &
				\includegraphics[width=.145\textwidth,height=2.1cm]{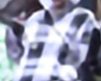} \hspace{-1em} &
				\includegraphics[width=.145\textwidth,height=2.1cm]{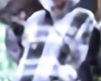} \hspace{-1em} &
				\includegraphics[width=.145\textwidth,height=2.1cm]{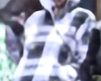}\vspace{0.42em}
				\\
				{\small (a) RCAN\hspace{-1em}}
			    &{\small (b) VESPCN \hspace{-1em}}
			    &{\small (c) TOFlow\hspace{-1em}}
		    	&{\small (d) FRVSR\hspace{-1em}}\vspace{0.42em}
				\\
			
				\includegraphics[width=.145\textwidth,height=2.1cm]{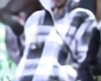} \hspace{-1em} &
				\includegraphics[width=.145\textwidth,height=2.1cm]{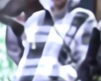} \hspace{-1em} &
				\includegraphics[width=.145\textwidth,height=2.1cm]{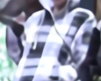} \hspace{-1em} &
				\includegraphics[width=.145\textwidth,height=2.1cm]{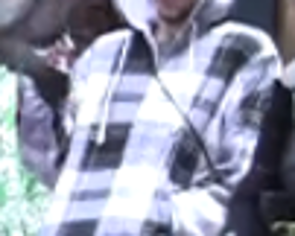}\vspace{0.5em}
				\\
				{\small (e) DUF\hspace{-1em}}
			    &{\small (f) RBPN\hspace{-1em}}
			    &{\small (g) Ours \hspace{-1em}}
			    &{\small (h) GT\hspace{-1em}}\vspace{0.5em}\\

			\end{tabular}
		\end{adjustbox}
		\vspace{0.5mm}
		\\
        \end{tabular}

		\caption{Visual results on Vid4 for $4\times$ scaling factor. Zoom in to see better visualization.}
		\label{fig:result_vid4}
	\end{center}\vspace{-1em}
\end{figure*}

\subsection{SR Reconstruction Module}
The output of the non-local attention module $F_{i}^{a'}$ is aggregated with the reference feature $F_{t}$ through a feature fusion layer, and then fed into the following SR reconstruction module. The SR reconstruction module mainly consists of stacked residual in residual dense blocks (RRDB) and a global skip connection. 

The structure of RRDB can be seen in Fig.\ref{figure:rrdb}. It combines multi-level residual network and dense connections. Benefiting from them, RRDBs can make full use of hierarchical features from input frames and obtain better restoration quality.  More details about RRDB can be found in \cite{wang2018esrgan}. The global skip connection transfers the shallow features of the reference frame to the end of the network, making the reconstruction module focus on learning residual features from the neighboring frames. It can well keep spatial information of the input LR reference frame and ensure the input frame and the corresponding super-resolved one have more structural similarity. Finally, a high-resolution reference frame is produced by a sub-pixel upsampling layer and a reconstruction layer.

\section{Experiments}

\subsection{Training Datasets and Details} 
\noindent \textbf{Datasets }
To train high-performance VSR networks, a large video dataset is required. Xue~\cite{xue2017video} et al. collected videos from Vimeo and released a VSR dataset vimeo-90k after processing. The dataset contains 64612 training samples with various and complex real-world motions. Each sample contains seven consecutive frames with a fixed resolution of $448 \times 256$. We use the vimeo-90k dataset as our training dataset. To generate LR images, we downscale the HR images $4 \times$ with MATLAB imresize function, which first blurs the input frames using cubic filters and then downsamples them using bicubic interpolation.

\begin{table*}[t!]
\scriptsize
  \begin{center}
\caption{Quantitative comparison of state-of-the-art SR algorithms on Vid4 for $4 \times$. {\color{red}Red} indicates the best and {\color{blue}blue} indicates the second best performance (PSNR/SSIM). In the evaluation, the first and last two frames are not included and we do not crop any border pixels except DUF. Eight pixels near image boundary are cropped for DUF.}
\label{tab:Vid4}
\begin{tabular}{l|c||c|c||c|c|c|c|c|c}
    \hline
      &Flow &Bicubic & RCAN~\cite{zhang2018image}& VESPCN~\cite{caballero2017real}& TOFlow~\cite{xue2017video}&FRVSR~\cite{sajjadi2018frame}&DUF~\cite{jo2018deep} & RBPN~\cite{haris2019recurrent}&DNLN(Ours)\\
      Clip Name &Magnitude&(1 Frame) & (1 Frame) & (3 Frames) & (7 Frames)  & (recurrent)&(7 Frames) & (7 Frames) & (7 Frames)  \\
      \hline
      Calendar	&1.14&20.39 / 0.5720 	&22.31 / 0.7248 	& -		&22.44 / 0.7290 	& -		&{\color{blue}24.07 / 0.8123} 	&23.95 / 0.8076 	&{\color{red}24.12  / 0.8141} \\
      City	    &1.63&25.17 / 0.6024 	&26.07 / 0.6938 	& -		&26.75 / 0.7368 	& -		&{\color{red}28.32 / 0.8333} 	&27.74 / 0.8051 	&{\color{blue}27.90 / 0.8111} \\
      Foliage	    &1.48&23.47 / 0.5666 	&24.69 / 0.6628 	& -		&25.24 / 0.7065 	& -		&{\color{red}26.41 / 0.7713} 	&26.21 / 0.7578 	&{\color{blue}26.28 / 0.7607} \\
      Walk	    &1.44&26.11 / 0.7977 	&28.64 / 0.8718 	& -		&29.03 / 0.8777 	& -		&30.63 / {\color{red}0.9144} 	&{\color{blue}30.70} / 0.9111 	&{\color{red}30.85} / {\color{blue}0.9129} \\
      \hline
      Average	    &1.42&23.79 / 0.6347 	&25.43 / 0.7383 &25.35 / 0.7557 &25.86 / 0.7625 &26.69 / 0.822 &{\color{red}27.36 / 0.8328} 	&27.15 / 0.8204 	&{\color{blue}27.29 / 0.8247} \\
    \hline
\end{tabular}
  \end{center}
\end{table*}

\begin{table*}[t!]
\scriptsize
  \begin{center}
\caption{Quantitative comparison of state-of-the-art SR algorithms on SPMCS-11 for $4 \times$.}
\label{tab:SPMCS-11}
\begin{tabular}{l|c||c|c||c|c|c|c}
    \hline
      &Flow &Bicubic & RCAN~\cite{zhang2018image}& TOFlow~\cite{xue2017video}&DUF~\cite{jo2018deep} & RBPN~\cite{haris2019recurrent}&DNLN(Ours)\\
      Clip Name &Magnitude&(1 Frame) &  (1 Frame) & (7 Frames) &(7 Frames) & (7 Frames) & (7 Frames)  \\
      \hline
      car05$\_$001 &6.21&27.75 / 0.7825 	&29.86 / 0.8484 	&30.10 / 0.8626 	&30.79 / 0.8707 	&{\color{blue}31.95} / {\color{red}0.9021} 	&{\color{red}31.96} / {\color{blue}0.9011} \\
      hdclub$\_$003$\_$001 &0.70&19.42 / 0.4863 	&20.41 / 0.6096 	&20.86 / 0.6523 	&{\color{blue}22.05} / {\color{red}0.7438} 	&21.91 / 0.7257 	&{\color{red}22.15} / {\color{blue}0.7366} \\
      hitachi$\_$isee5$\_$001 &3.01&19.61 / 0.5938 	&23.71 / 0.8369 	&22.88 / 0.8044 	&25.77 / 0.8929 	&{\color{blue}26.30 / 0.9049} 	&{\color{red}26.60 / 0.9080} \\
      hk004$\_$001 &0.49&28.54 / 0.8003 	 &31.68 / 0.8631 	&30.89 / 0.8654 	&32.98 / 0.8988 	&{\color{blue}33.38 / 0.9016} 	&{\color{red}33.46 / 0.9041} \\
      HKVTG$\_$004 &0.11&27.46 / 0.6831 	&28.81 / 0.7649 	&28.49 / 0.7487 	&29.16 / 0.7860 	&{\color{blue}29.51} / {\color{red}0.7979} 	&{\color{red}29.53} / {\color{blue}0.7976} \\
      jvc$\_$009$\_$001 &1.24&25.40 / 0.7558 	&28.31 / 0.8717 	&27.85 / 0.8542 	&29.18 / 0.8961 	&{\color{blue}30.06 / 0.9105} 	&{\color{red}30.65 / 0.9205} \\
      NYVTG$\_$006 &0.10&28.45 / 0.8014 	&31.01 / 0.8859 	&30.12 / 0.8603 	&32.30 / 0.9090 	&{\color{blue}33.22 / 0.9231} 	&{\color{red}33.35 / 0.9254} \\
      PRVTG$\_$012 &0.12&25.63 / 0.7136 	&26.56 / 0.7806 	&26.62 / 0.7788 	&27.39 / 0.8166 	&{\color{blue}27.60 / 0.8242} 	&{\color{red}27.68 / 0.8260} \\
      RMVTG$\_$011 &0.18&23.96 / 0.6573 	&26.02 / 0.7569 	&25.89 / 0.7500 	&27.56 / 0.8113 	&{\color{blue}27.63 / 0.8170} 	&{\color{red}27.75 / 0.8199} \\
      veni3$\_$011 &0.36&29.47 / 0.8979 	&34.58 / 0.9629 	&32.85 / 0.9536 	&34.63 / 0.9677 	&{\color{red}36.61} / {\color{blue}0.9735} 	&{\color{blue}36.33} / {\color{red}0.9739} \\
      veni5$\_$015 &0.36&27.41 / 0.8483 	&31.04 / 0.9262 	&30.03 / 0.9118 	&31.88 / 0.9371 	&{\color{blue}32.37 / 0.9409} 	&{\color{red}33.04 / 0.9466} \\
      \hline
      Average &1.17&25.73 / 0.7291 	&28.36 / 0.8279 	&27.87 / 0.8220 	&29.43 / 0.8664 	&{\color{blue}30.05 / 0.8747} 	&{\color{red}30.23 / 0.8782} \\
    \hline
\end{tabular}
  \end{center}
\end{table*}

\begin{table}[t!]
\scriptsize
  \begin{center}
\caption{Quantitative comparison of state-of-the-art SR algorithms on Vimeo-90K-T for $4 \times$.}
\label{tab:Vimeo-90K-T}
\setlength{\tabcolsep}{1.4mm}{
\begin{tabular}{*1l||*1c|*1c|*1c|*1c}
\hline     
Method  & Slow & Medium & Fast & Average  \\
\hline
Bicubic	    &29.34 / 0.8330 	&31.29 / 0.8708 	&34.07 / 0.9050 	&31.32 / 0.8684\\
RCAN	   ~\cite{zhang2018image}  &32.93 /	0.9032 	&35.35 / 0.9268 	&38.47 / 0.9456 	&35.34 / 0.9249\\
\hline
TOFlow	   ~\cite{xue2017video}  &32.15 / 0.8900 	&35.01 / 0.9254 	&37.70 / 0.9430 	&34.84 / 0.9209\\
DUF	       ~\cite{jo2018deep}  &33.41 /	0.9110 	&36.71 / 0.9446 	&38.87 / 0.9510 	&36.37 / 0.9386\\
RBPN	   ~\cite{haris2019recurrent}  &{\color{blue}34.26 / 0.9222} 	&{\color{blue}37.39 / 0.9494} 	&{\color{blue}40.16 / 0.9611} 	&{\color{blue}37.18 / 0.9456}\\
DNLN(ours) 	&{\color{red}34.47 / 0.9246} 	&{\color{red}37.59 / 0.9510} 	&{\color{red}40.35 / 0.9621} 	&{\color{red}37.38 / 0.9473}\\
\hline
\# of clips &1616 &4983 &1225 &7824\\
Flow Mag. &0.6 &2.5 &8.3 &3.0\\
\hline
\end{tabular}}
  \end{center}
\end{table}

\noindent \textbf{Training Details }
In our network, the convolutional layers have 64 filters and their kernel sizes are set to $3 \times 3$, if not specified otherwise. In the feature extraction module, we utilize 5 residual blocks to extract shallow features. Then the alignment module adopts 5 deformable convolutions to perform feature alignment. The dilated convolutions in HFFB have $3 \times 3$ kernels and 32 filters. In the non-local attention module, the first three $1 \times 1$ convolutions have 32 filters and the last $1 \times 1$ convolution has 64 filters. Finally, in the reconstruction module, we use 23 RRDBs and set the number of growth channels to 32. 

In the training process, we perform data augmentation by doing horizontal or vertical flipping, $90 ^ { \circ }$ rotation and random cropping of the images. The batch size is set to 8. The network takes seven consecutive frames as inputs, and LR patches with the size of $50 \times 50$ are extracted for training. Our model is trained by Adam optimizer~\cite{kingma2014adam} with $\beta_{1}=0.9$, $\beta_{2}=0.999$, and $\epsilon=10^{-8}$. The initial learning rate is $10^{-4}$ before 70 epochs and later decreases to half every 20 epochs. All experiments were conducted on two NVIDIA RTX 2080 GPUs using PyTorch 1.0~\cite{paszke2017automatic}. We train the network end-to-end by minimizing L1 loss between the predicted frame and the ground truth HR frame. And later we employ the L2 loss to finetune the model, which could result in better performance.


\begin{figure*}[htb]
  \begin{center}
    \begin{tabular}[c]{cccccc}
    
       \includegraphics[width=.198\textwidth]{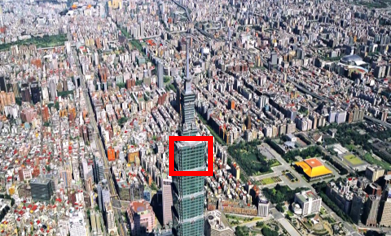}\hspace{-1em}
       &
      \includegraphics[width=.15\textwidth]{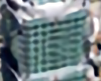}\hspace{-1em}
      &
      \includegraphics[width=.15\textwidth]{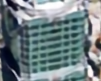}\hspace{-1em}
      &
      \includegraphics[width=.15\textwidth]{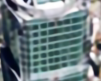}\hspace{-1em}
      &
      \includegraphics[width=.15\textwidth]{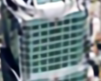}\hspace{-1em}
       &
       \includegraphics[width=.15\textwidth]{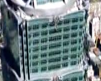}\hspace{-1em}\\
        
    \includegraphics[width=.198\textwidth]{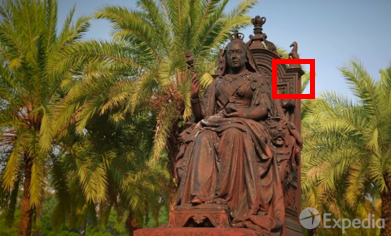}\hspace{-1em}
    &
    \includegraphics[width=.15\textwidth]{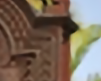}\hspace{-1em}
    &
    \includegraphics[width=.15\textwidth]{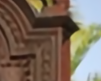}\hspace{-1em}
    &
    \includegraphics[width=.15\textwidth]{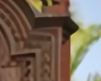}\hspace{-1em}
    &
    \includegraphics[width=.15\textwidth]{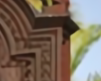}\hspace{-1em}
    &
    \includegraphics[width=.15\textwidth]{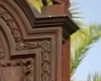}\hspace{-1em}\\
    
    \includegraphics[width=.198\textwidth]{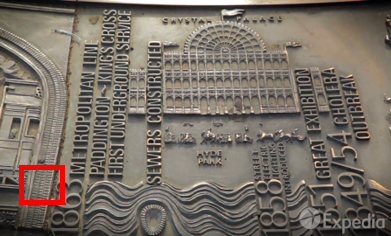}\hspace{-1em}
    &
    \includegraphics[width=.15\textwidth]{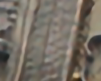}\hspace{-1em}
    &
    \includegraphics[width=.15\textwidth]{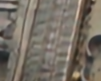}\hspace{-1em}
    &
    \includegraphics[width=.15\textwidth]{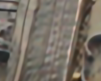}\hspace{-1em}
    &
    \includegraphics[width=.15\textwidth]{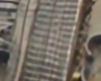}\hspace{-1em}
    &
    \includegraphics[width=.15\textwidth]{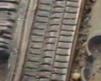}\hspace{-1em}\\
    
    \includegraphics[width=.198\textwidth]{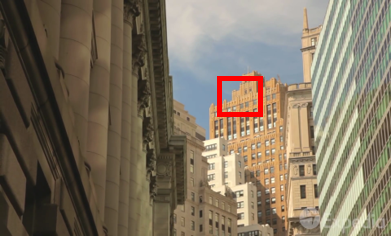}\hspace{-1em}
    &
    \includegraphics[width=.15\textwidth]{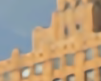}\hspace{-1em}
    &
    \includegraphics[width=.15\textwidth]{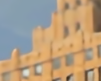}\hspace{-1em}
    &
    \includegraphics[width=.15\textwidth]{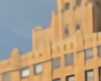}\hspace{-1em}
    &
    \includegraphics[width=.15\textwidth]{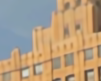}\hspace{-1em}
    &
    \includegraphics[width=.15\textwidth]{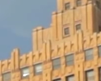}\hspace{-1em}\\
    
    \includegraphics[width=.198\textwidth]{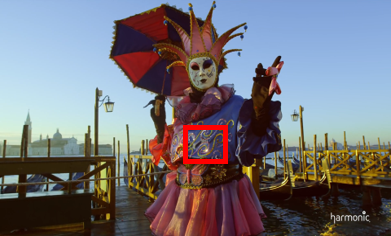}\hspace{-1em}
    &
    \includegraphics[width=.15\textwidth]{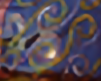}\hspace{-1em}
    &
    \includegraphics[width=.15\textwidth]{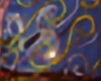}\hspace{-1em}
    &
    \includegraphics[width=.15\textwidth]{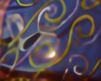}\hspace{-1em}
    &
    \includegraphics[width=.15\textwidth]{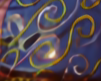}\hspace{-1em}
    &
    \includegraphics[width=.15\textwidth]{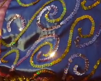}\hspace{-1em}\\

      &{\small (a) TOFlow\hspace{-1em}}
      &{\small (b) DUF\hspace{-1em}}
      &{\small (c) RBPN\hspace{-1em}}
      &{\small (d) Ours\hspace{-1em}}
      &{\small (e) GT \vspace{0.2em}}\\
    \end{tabular}
    \caption{Visual results on SPMCS for $4\times$ scaling factor. Zoom in to see better visualization.}
    \label{fig:result_spmcs}
  \end{center}\vspace{-1em}
\end{figure*}


\begin{figure*}[!t]
  \begin{center}
    \begin{tabular}[c]{cccccc}
      \includegraphics[width=.15\textwidth]{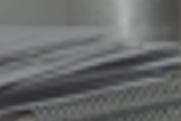}\hspace{-0.9em}
      &
      \includegraphics[width=.15\textwidth]{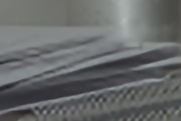}\hspace{-0.9em}
      &
      \includegraphics[width=.15\textwidth]{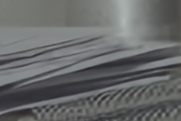}\hspace{-0.9em}
      &      
      \includegraphics[width=.15\textwidth]{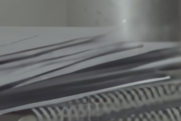}\hspace{-0.9em}
      &
      \includegraphics[width=.15\textwidth]{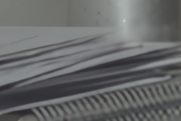}\hspace{-0.9em}
        &
        \includegraphics[width=.15\textwidth]{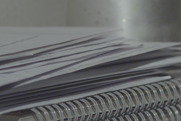}\hspace{-0.9em}\\
        
    \includegraphics[width=.15\textwidth]{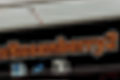}\hspace{-0.9em}
      &
      \includegraphics[width=.15\textwidth]{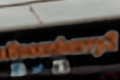}\hspace{-0.9em}
      &
      \includegraphics[width=.15\textwidth]{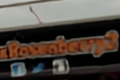}\hspace{-0.9em}
      &      
      \includegraphics[width=.15\textwidth]{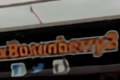}\hspace{-0.9em}
      &
      \includegraphics[width=.15\textwidth]{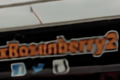}\hspace{-0.9em}
        &
        \includegraphics[width=.15\textwidth]{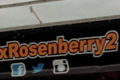}\hspace{-0.9em}\\
    
    \includegraphics[width=.15\textwidth]{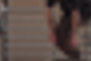}\hspace{-0.9em}
    &
    \includegraphics[width=.15\textwidth]{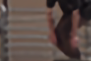}\hspace{-0.9em}
    &
    \includegraphics[width=.15\textwidth]{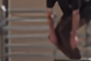}\hspace{-0.9em}
    &      
    \includegraphics[width=.15\textwidth]{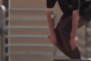}\hspace{-0.9em}
    &
    \includegraphics[width=.15\textwidth]{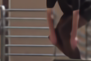}\hspace{-0.9em}
    &
    \includegraphics[width=.15\textwidth]{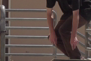}\hspace{-0.9em}\\
    
    \includegraphics[width=.15\textwidth]{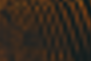}\hspace{-0.9em}
    &
    \includegraphics[width=.15\textwidth]{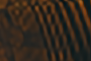}\hspace{-0.9em}
    &
    \includegraphics[width=.15\textwidth]{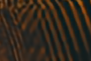}\hspace{-0.9em}
    &      
    \includegraphics[width=.15\textwidth]{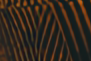}\hspace{-0.9em}
    &
    \includegraphics[width=.15\textwidth]{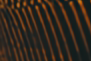}\hspace{-0.9em}
    &
    \includegraphics[width=.15\textwidth]{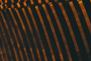}\hspace{-0.9em}\\

    \includegraphics[width=.15\textwidth]{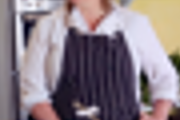}\hspace{-0.9em}
      &
      \includegraphics[width=.15\textwidth]{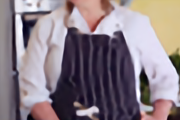}\hspace{-0.9em}
      &
      \includegraphics[width=.15\textwidth]{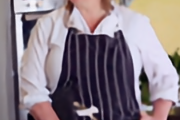}\hspace{-0.9em}
      &      
      \includegraphics[width=.15\textwidth]{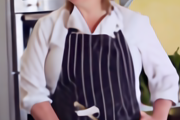}\hspace{-0.9em}
      &
      \includegraphics[width=.15\textwidth]{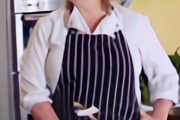}\hspace{-0.9em}
        &
        \includegraphics[width=.15\textwidth]{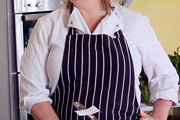}\hspace{-0.9em}\\

      {\small (a) Bicubic\hspace{-0.9em}}
      &{\small (b) TOFlow\hspace{-0.9em}}
      &{\small (c) DUF\hspace{-0.9em}}
      &{\small (d) RBPN\hspace{-0.9em}}
      &{\small (e) Ours\hspace{-0.9em}}
      &{\small (f) GT \vspace{0.3em}}\\
    \end{tabular}
    \caption{Visual results on Vimeo-90K-T for $4\times$ scaling factor. Zoom in to see better visualization.}
    \label{fig:result_vimeo}
  \end{center}\vspace{-1.5em}
\end{figure*}

\begin{figure*}[htb]
	\begin{center}
		\begin{tabular}[c]{ccccccc}
			\includegraphics[width=.13\textwidth]{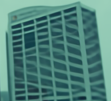}\hspace{-0.9em}
			&
			\includegraphics[width=.13\textwidth]{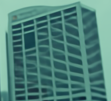}\hspace{-0.9em}
			&
			\includegraphics[width=.13\textwidth]{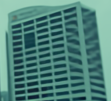}\hspace{-0.9em}
			&      
			\includegraphics[width=.13\textwidth]{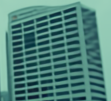}\hspace{-0.9em}
			&
			\includegraphics[width=.13\textwidth]{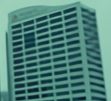}\hspace{-0.9em}
			&
			\includegraphics[width=.13\textwidth]{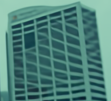}\hspace{-0.9em}
			&
			\includegraphics[width=.13\textwidth]{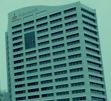}\hspace{-0.9em}\\

			{\small (a) 1dconv\hspace{-0.9em}} 
			&{\small (b) 2dconv\hspace{-0.9em}}
			&{\small (c) 3dconv\hspace{-0.9em}}
			&{\small (d) 4dconv\hspace{-0.9em}}
			&{\small (e) 5dconv\hspace{-0.9em}}
			&{\small (f) w/o HFFB  \hspace{-0.9em}}
			&{\small (g) GT \vspace{0.3em}}\\

		\end{tabular}
		\caption{Qualitative results of ablation on alignment module.}
		\label{fig:result_ablation}
	\end{center}\vspace{-1.5em}
\end{figure*}

\subsection{Comparison with the State-of-the-art Methods} 
We compare our DNLN with several state-of-the-art SISR and VSR methods: RCAN~\cite{zhang2018image}, VESPCN~\cite{caballero2017real}, TOFlow~\cite{xue2017video}, FRVSR~\cite{sajjadi2018frame}, DUF~\cite{jo2018deep} and RBPN~\cite{haris2019recurrent}. Note that most previous methods are trained with different datasets and we just compare with the results they provided. The SR results are evaluated with PSNR and SSIM~\cite{wang2004image} quantitatively on Y channel (i.e., luminance) of transformed YCbCr space. In the evaluation, the first and last two frames are not included and we do not crop any border pixels except DUF~\cite{jo2018deep}. Eight pixels near image boundary are cropped for DUF due to its severe boundary effects. 

We evaluated our models on three datasets: Vid4~\cite{liu2013bayesian}, SPMCS~\cite{tao2017detail}, and Vimeo-90K-T~\cite{xue2017video} with average flow magnitude (pixel/frame) provided in \cite{haris2019recurrent}. Vid4 is a commonly used dataset which contains four video sequences: calendar, city, foliage and walk. However, we can observe that Vid4 has limited inter-frame motion and there exists artifacts on its ground-truth frames. SPMCS consists of higher quality video clips with various motions and diverse scenes. Vimeo-90K-T is a much larger dataset. It contains a wide range of flow magnitude between frames which can well judge the performance of the VSR methods. 

Table~\ref{tab:Vid4} shows the quantitative results on Vid4. Our model outperforms the optical-flow-based methods which demonstrates the effectiveness of our optical flow free alignment module. Qualitative results are shown in Fig.\ref{fig:result_vid4}. We mark out the positions which display obvious distinctions among different methods. For the date in “Calendar” clip, most compared methods produce images with blurring artifacts, while our method achieves a better result and alleviates the artifacts. In the “Walk” clip, the existing methods blur the rope and clothing together, only DNLN can clearly distinguish these two parts and restore the pattern closest to the ground truth frame. 

In comparison to Vid4, SPMCS contains more high frequency information with higher resolution, which requires the superb recovery abilities of algorithms. Results on SPMCS are shown in Table~\ref{tab:SPMCS-11}. DNLN achieves the best results and outperforms other methods by a large margin on PSNR, which proves the superiority of our model. Visual comparisons are depicted in Fig.\ref{fig:result_spmcs}. Due to the abundance of textures, most methods cannot fully recover the frames and obviously produce blurring artifacts. Although DUF and RBPN could reproduce part of the HR patterns, it is obvious that our DNLN is the unique approach to restore the abundant details and clean edges. Such visual comparisons demonstrate that our network can extract more sophisticated features from LR space with the proposed modules.

Table~\ref{tab:Vimeo-90K-T} presents the quantitative outcomes of Vimeo-90K-T. As suggested in~\cite{haris2019recurrent}, we classified the video clips into three groups (e.g. slow, medium and fast) according to the motion velocity. While the motion velocity increases, video frames with larger motion amplitude will contain more useful temporal information but also make the recovery more challenging. Our DNLN ensures optimal performance on all three groups, surpassing RBPN by 0.21 dB, 0.20 dB and 0.19 dB on PSNR, respectively. Since the flow magnitude of Vimeo-90K-T in fast group is higher than Vid4 and SPMCS, the content between video frames varies greatly, which reflects that DNLN could take full advantage of temporal information among multiple frames. The qualitative evaluations are shown in Fig.\ref{fig:result_vimeo}. For the railing texture in third row, only our method restores the correct and clear pattern while others suffer from varying degrees of blurring. In some cases, even the SR frames recovered by different methods have the same sharp edges, our DNLN is more accurate and faithful to the ground truth. Such as the images in fourth row, the results restored by RBPN and DNLN are equally clear, while the former produces the stripes with wrong directions. 

\subsection{Model Size and Running Time Analyses} 
Table~\ref{tab:model_size} shows comparisons about model size and and running time of the methods. The running time is test with input size $112\times64$. Our DNLN has the largest model size but also achieves the best performance. Here, we adopt a smaller model S-DNLN with only 3 deformable convolutions in alignment module and 14 RRDBs in reconstruction module. We can see that S-DNLN has a comparable number of parameters and running time with RBPN. Nevertheless, it can still get a better result which further validates the effectiveness of our proposed modules.

\begin{table}[htb]
	\scriptsize
	\begin{center}
		\caption{Number of parameters and time cost on Vimeo-90K-T for $4\times$.}
		\label{tab:model_size}
		\begin{tabular}{c|c|c|c|c|c|c}
			\hline
			Methods & RCAN & TOFlow & DUF & RBPN & S-DNLN & DNLN\\
			\hline
			Parameter (M) &15.59 &1.41 &5.82 &12.77 &12.39 &19.74\\
			\hline
			Time (s)  &0.051 &0.126 &0.175 &0.086 &0.095 &0.119\\
			\hline
			PSNR  &35.34 &34.84 &36.37 &37.18 &37.23 &37.38\\
			\hline
		\end{tabular}
	\end{center}
\end{table}

\section{Ablation Study} 
\begin{figure}[t!]
	\centering
	\includegraphics[width=\linewidth]{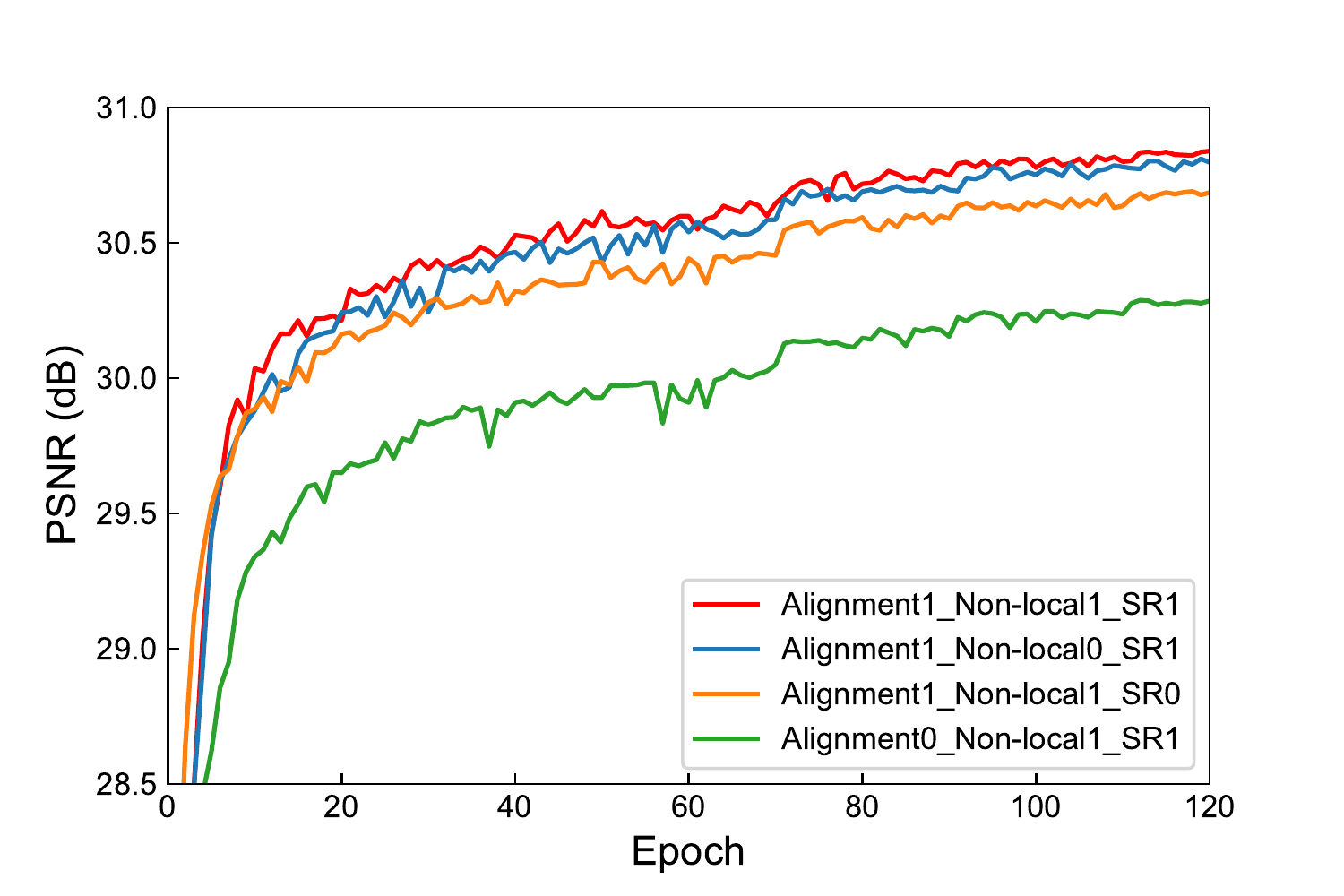}
	\caption{Convergence analysis on the three modules of proposed network. The curves for each combination are based on the PSNR with scaling factor $4 \times$.}
	\label{figure:convergence analysis}
\end{figure}

To further investigate the proposed method, we conducted ablation experiments by removing the main components of our network. The results are shown in Table~\ref{tab:ablation_dnln}. First, we remove the alignment module, thus the shallow features would be directly fed into the following network without warping. The PSNR of the results on Vimeo-90K-T is relatively low, which indicates that the alignment operation is crucial for utilizing the inter-frame information. Second, we remove the non-local attention module and the performance decreases a lot. Third, we replace the RRDBs by simply stacking common residual blocks and it also harms the performance. We visualize the convergence process of these combinations in Fig.\ref{figure:convergence analysis}. The results demonstrate the effectiveness and benefits of our proposed three modules.

\begin{table}[H]
	\scriptsize
	\begin{center}
		\caption{Ablation study of proposed network on Vimeo-90K-T for $4 \times$.}
		\label{tab:ablation_dnln}
		\begin{tabular}{ccc|c|c}
			\hline
			\begin{tabular}[c]{@{}c@{}}Alignment \\ module  \end{tabular} &
			\begin{tabular}[c]{@{}c@{}}Non-local \\ module  \end{tabular} & 
			\begin{tabular}[c]{@{}c@{}c@{}}SR    \\ module  \end{tabular} &  PSNR & SSIM   \\ 
			\hline
			& \Checkmark & \Checkmark &  36.82 & 0.9418   \\
			\Checkmark &            & \Checkmark &  37.34 & 0.9471   \\
			\Checkmark & \Checkmark &            &  37.25 & 0.9462   \\
			\Checkmark & \Checkmark & \Checkmark & \color{red}37.38 & \color{red}0.9473   \\
			\hline
		\end{tabular}
	\end{center}
\end{table}

From the ablation experiments above, we can observe that the network performance would be significantly affected by the alignment preprocessing. So we further validated the impact of deformable convolutions on the reconstruction capability. As shown in Table~\ref{tab:ablation_alignment}, with only one deformable convolution, the PSNR value can improve greatly. It demonstrates the importance of alignment operations for making efficient use of the neighboring frames. As the number of deformable convolutions increases, the network gains a better performance. The visual comparisons are shown in Fig.\ref{fig:result_ablation}. From left to right, the network alleviates the blurring artifacts of the office building and recovers more accurate details. In addition, we replaced the HFFB used in deformable convolutions with a $3 \times 3$ convolution layer. The performance of network decreases by roughly 0.29 dB. It proves that by enlarging the receptive field, the deformable convolution can more effectively cope with complex and large motions.

\begin{table}[H]
	\scriptsize
	\begin{center}
		\caption{Ablation study on alignment module.}
		\label{tab:ablation_alignment}
		\begin{tabular}{c|c|c}
			\hline
			Model & PSNR & SSIM    \\
			\hline
			w/o deform &36.81 & 0.9417 \\
			1dconv	   &37.09 & 0.9446 \\
			2dconv	   &37.19 & 0.9455 \\
			3dconv	   &37.33 & 0.9469 \\
			4dconv	   &37.36 & 0.9471 \\
			5dconv	   &\color{red}37.38 & \color{red}0.9473 \\
			5dconv, w/o HFFB &37.09 & 0.9447 \\
			\hline
		\end{tabular}
	\end{center}
\end{table}

In order to study the influence of inter-frame information on the recovery results, we leveraged different number of frames to train our network. From Table~\ref{tab:ablation_frames}, we can observe that there is a significant improvement in DNLN when switching from 3 frames to 5 frames, and the performance of DNLN/5 is even better than RBPN which uses 7 frames. When further switching to 7 frames, we can still get a better result but the improvement becomes minor.

\begin{table}[H]
	\scriptsize
	\begin{center}
		\caption{Experimental results with a different number of input frames.}
		\label{tab:ablation_frames}
		\begin{tabular}{c|c|c|c}
			\hline
			Input & 3 frames & 5 frames & 7 frames    \\
			\hline
			PSNR / SSIM &37.06 / 0.9435 &37.29 / 0.9463 &{\color{red}37.38 / 0.9473}\\
			\hline
		\end{tabular}
	\end{center}
\end{table}

\section{Conclusion}
In this paper, we propose a novel deformable non-local network (DNLN), which is a non-flow-based method for effective video super-resolution. To deal with complicated and large motion compensation, we introduce the deformable convolution with HFFB in our alignment module, which can well align the frames at the feature level. In addition, we adopt a non-local attention module to further extract complementary features from neighboring frames. By making full use of the temporal information, we finally restore a high quality video frame through a reconstruction module. Extensive experiments on benchmark datasets illustrate the effectiveness of our DNLN in video super-resolution.


%

\ifCLASSOPTIONcaptionsoff
  \newpage
\fi



%

\bibliographystyle{IEEEtran}
\bibliography{refs}

%




\end{document}